\documentclass[twocolumn,showpacs,preprintnumbers,amsmath,amssymb]{revtex4}

\usepackage{graphicx}
\usepackage{dcolumn}
\usepackage{bm}
\usepackage{psfrag}
\usepackage{color}
\usepackage{subfigure}
\usepackage{dsfont}

\newcommand{\bea}{\begin{eqnarray}}
\newcommand{\eea}{\end{eqnarray}}

\newcommand{\ee}  {{\text{e}}}
\newcommand{\vect}[1]{\mathbf{#1}}

\newcommand{\op}[1]  {\hat{#1}}
\newcommand{\OP}[1]  {\hat{\cal #1}}

\newcommand{\HH}{{\hat{\cal H}}}

\newcommand{\T}{{}_{[T]}}
\newcommand{\nuc}{{}_{\text{nuc}}}

\newcommand{\Tr}       {\text{Tr}}
\newcommand{\betap}    {\frac{\beta}{P}}

\newcommand{\pone}{{}^{(1)}}
\newcommand{\ptwo}{{}^{(2)}}

\newcommand{\pp}{{}^{(p)}}
\newcommand{\ppp}{{}^{(p+1)}}

\newcommand{\prot}{{}$^1$H}
\newcommand{\deut}{{}$^2$H}

\begin{document}
\title{The isotope-effect in the phase transition of KDP:
New insights from ab-initio path-integral simulations.}

\author
{V. Srinivasan$^{(1)}$, R. Car$^{(1,2)}$, and D. Sebastiani$^{(3,4)}$}
\affiliation{(1) Department of Chemistry, 
Princeton University, Princeton, NJ 08544}
\affiliation{(2) Department of Physics,
Princeton University, Princeton, NJ 08544}
\affiliation{(3) Max-Planck Institut f\"{u}r Polymerforschung, 
10 Ackermannweg, D 55128 Mainz, Germany}
\affiliation{(4) Department of Physics, FU Berlin, 
Arnimalle 14, 14195 Berlin, Germany}

\date{\today}

\begin{abstract}

We investigate the quantum-mechanical localization of \prot\ and \deut\ isotopes in the symmetric low-barrier hydrogen-bonds of potassium dihydrogen phosphate (KDP) crystals in the paraelectric phase. The spatial density distributions of these hydrogen atoms are suspected to be responsible for the surprisingly large isotope effect observed  for the ferroelectric phase transition in KDP. We employ {\it ab initio} path integral molecular dynamics simulations to obtain the nuclear real-space and momentum-space densities $n(\mathbf R)$ and $n(\mathbf k)$ of \prot\ and \deut, which are compared to experimental Neutron Compton Scattering data.

Our results suggest a qualitative difference in the nature of the paraelectic phase in KDP between the two isotopes. We are able to discriminate between real quantum delocalization and vibration-assisted hopping and thus provide evidence for two distinct mechanisms of the ferroelectric phase transition in this class of materials.

\end{abstract}

\pacs{61.12.Ex,    
73.40.Gk,    
78.70.Nx,    
77.80.-e,    
77.80.Bh,    
71.15.Ap,    
71.20.-b,    
71.23.-k,    
71.15.Pd,     
74.25.Fy, 72.15.Jf 
}

\maketitle
\newcommand{\trashA}{
SOME COMMENTS:
For a harmonic oscillator, $\Psi(r)\propto\exp - \frac 1 2 \frac{m\omega}{\hbar} x^2$

thus $n(r)\propto\exp - \frac 1 2 \frac {2 m \omega}{\hbar} x^2$

thus $\Psi(k)\propto\exp - \frac 1 2 \frac{\hbar}{m\omega} k^2$

thus $n(k)\propto\exp - \frac 1 2 \frac{ 2 \hbar} {m \omega} k^2$

For a free particle, $n(k)\propto\exp - \frac 1 2 \frac {\hbar^2}{m k T} k^2$

In conclusion: assume that $n(k)\propto\exp - \frac 1 2 \sigma^2 k^2$, then:
}

\begin{figure}[t!]
\begin{center}
\includegraphics[width=3.0in]{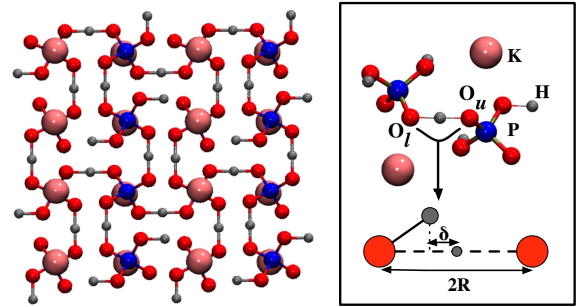}
\end{center}
\caption{Left : The paraelectric KDP unit cell along the polar $c$-axis. The H-bonds between phosphates lie along the $a$ and $b$-axes, respectively. Right :  two phosphate molecules with their potassium counterions which form the basic building block of a KDP unit cell. The ferroelectric polarization is caused by a symmetry breaking in the K -- P -- K distances along the $c$-axis. Bottom panel with a single OH$\cdots$O bond illustrates the proton-localization coordinate $\delta$ and the acceptor-donor distance $2R$ (see text). }
\label{OHO}
\end{figure}

Potassium dihydrogen phosphate (KDP) is the prototype of a wide class of hydrogen-bonded ferroelectrics that find use in non-linear optics applications~\cite{LinesGlass}. Its phosphate units are interconnected by a network of hydrogen-bonds (H-bonds). The ferroelectric to paraelectric phase transition in these ionic crystals is associated with a shift from asymmetric OH$\cdots$O H-bonds (fig.~\ref{OHO}) to a symmetric configuration. 

Earlier theories for the phase transition associated the symmetry of the paraelectric phase with a classical disordering of protons over the two donor sites in each H-bond~\cite{SlaterKDP,Takagi}. However, the observation of a very large isotope effect ($T_c$[\prot]=122K to $T_c$[\deut]=230K~\cite{Bantle}), strongly suggested the presence of quantum effects. As a result, the observed disorder was  attributed~\cite{Blinc} solely to \prot-tunneling between the two equivalent donor sites in each H-bond. The increased mass due to deuteration reduces the tunnel-splitting. As the asymmetric OH$\cdots$O bond is favored by electrostatic effects, it would then dominate the tunneling delocalization over a larger range of temperatures and lead to a higher $T_c$. 

This tunneling model assumed that the mass effect is dominant and that the effective potential energy surface (PES) for the \prot/\deut~motion remains the same. However, pressure-dependent studies on KDP~\cite{Nelmesrev} and related systems have shown~\cite{McMahon} that the tunneling hypothesis is insufficient to completely account for the observed increase in $T_c$. An additional  Ubbelohde~\cite{Ubbelohde}  effect  is assumed to contribute significantly to the increase in $T_c$. 

In this Letter we probe the nature of the paraelectric phases of protonated and deuterated KDP using quantum-mechanical density distributions of the hydrogen nuclei both in real and reciprocal (momentum-) space. We use a recently developed {\it ab initio} path-integral molecular dynamics (PIMD) method~\cite{morrone-pi, morrone-water}, which permits us to model simultaneously electronic and nuclear quantum effects at realistic temperatures and in the actual periodic crystal structure. We present quantitative results for the localization and delocalization of protons and deuterons, and thus distinguish between quantum-mechanical delocalization (tunneling) and classical hopping of localized particles, as a function of temperature. We show that this essential mechanistic difference between the ferroelectric transitions in protonated and deuterated KDP is key to understanding the huge isotope-effect observed in this sytem.

Solving the nuclear Schr\"odinger equation on a density-functional theory (DFT) based PES requires prohibitive computational resources except for very low-dimensional systems, i.e. with few nuclear degrees of freedom. Our calculations are based on the path integral technique, which represents a convenient way to map the nuclear quantum problem to a statistical ensemble average of a set of non-quantum replica of the same system. The general path-integral method is described in detail elsewhere~\cite{feynman-pi,pi-tuckerman-phd,tuckerman-pathintegral,voth-pi-1,voth-review, pi-ceperley-rmp}. 

\begin{figure*}[t!]
\subfigure[]{
\includegraphics[width=0.50\textwidth]{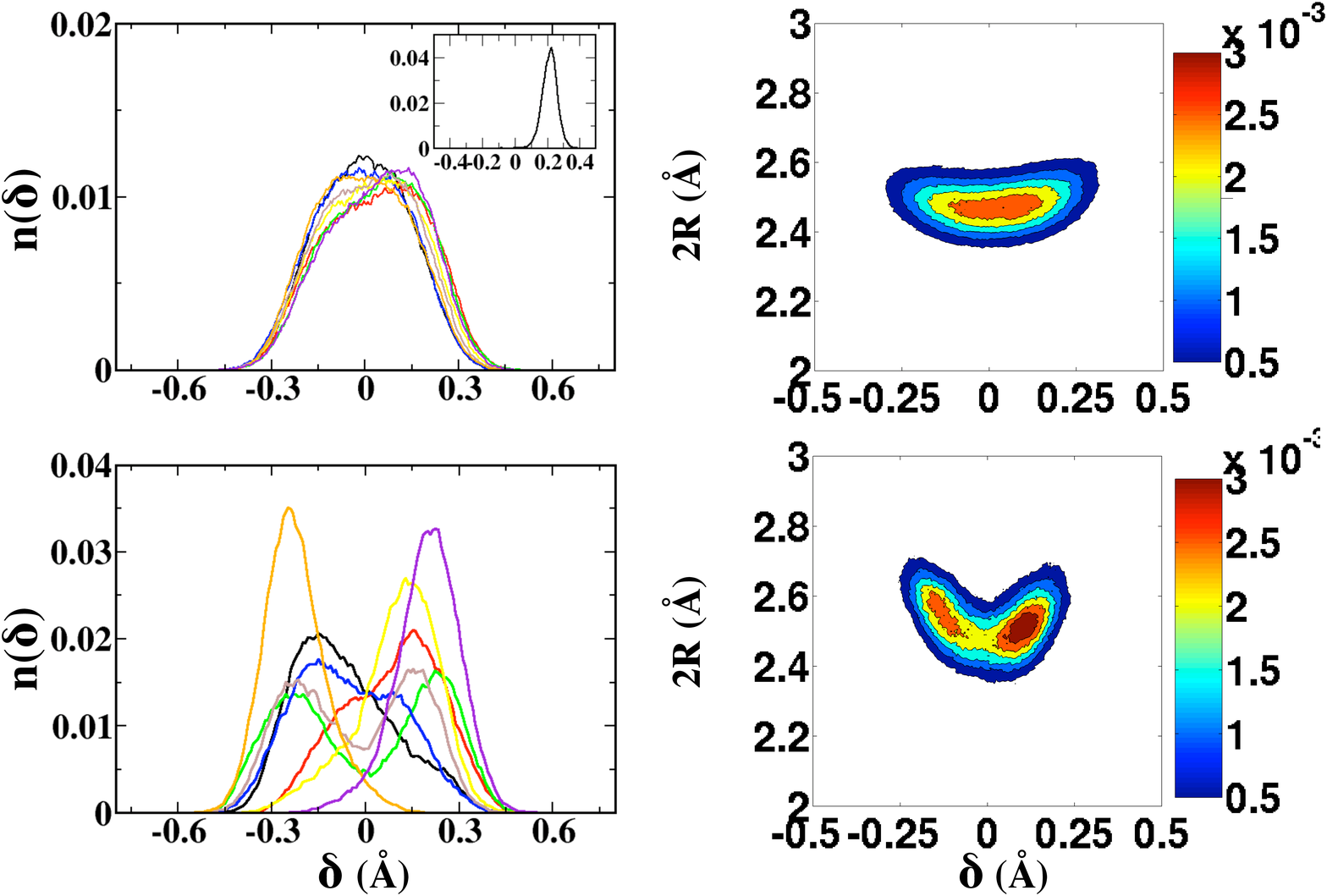}
}
\subfigure[]{
\includegraphics[width=0.45\textwidth]{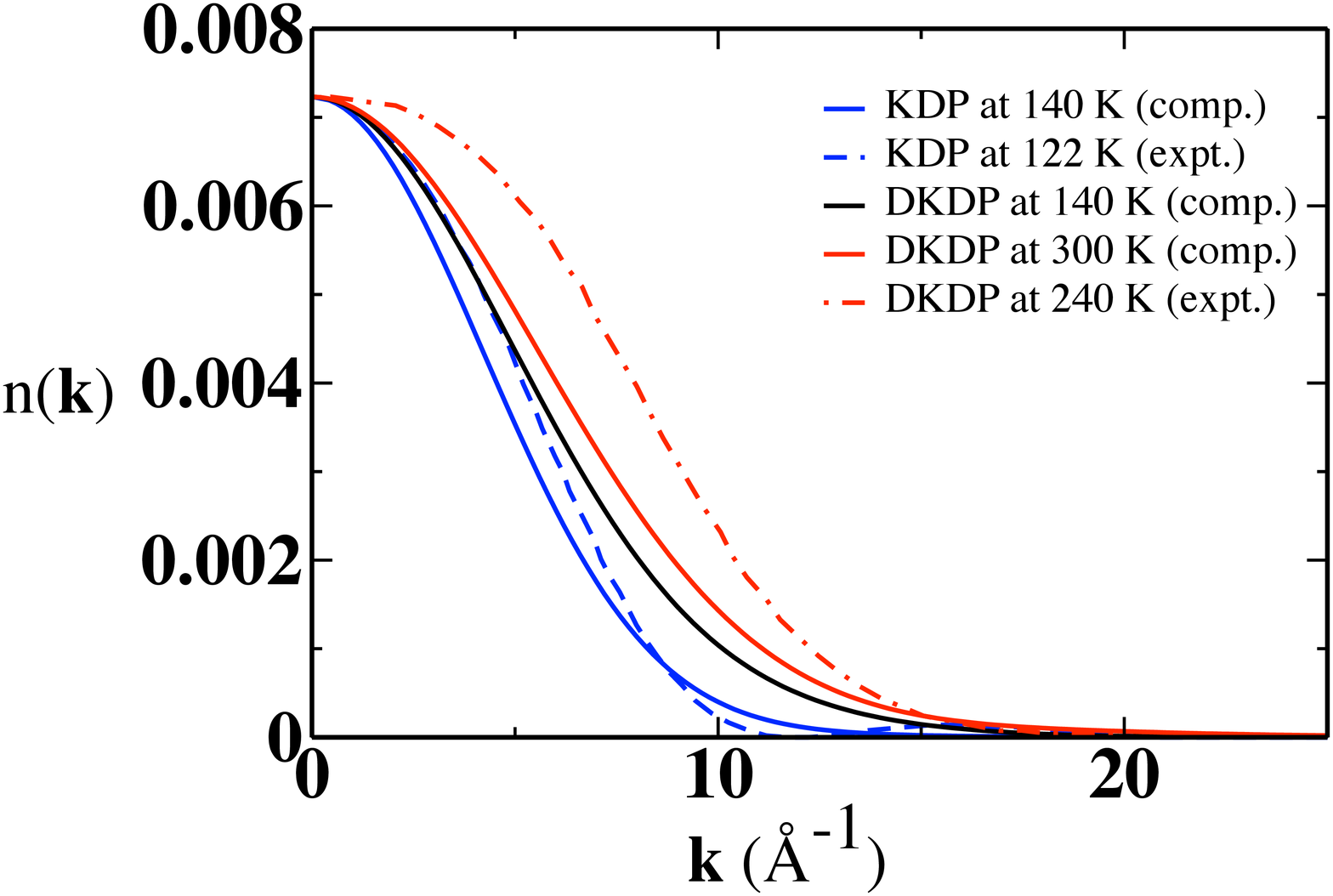}
}
\caption{ (a) Hydrogen localization functions HLF (left) and 2R-$\delta$ correlation (right) from \textbf{12}ps of {\it ab-initio} path-integral MD simulations for \prot-KDP at T=140K (top) and \deut-KDP at 300K (bottom) (paraelectric phases). The inset shows the HLF from a conventional MD simulation of the same system; (b) Average momentum distributions - computed and experimental (NCS)~\cite{ReiterKDP}.
\label{fig:PLF}}
\end{figure*}

In the common path integral formulation of a quantum system of $n$ nuclei, the canonical partition function $Z=\Tr\;\exp -\beta\HH = \Tr\;\op{\rho}\T$ is written as an integral over discretized paths $\lbrace \mathbf{R}^{(1)}\mathbf{R}^{(2)} \ldots \mathbf{R}^{(P)} ; \mathbf{R}^{(i)} \in {\Re}^{3n} \rbrace$ in coordinate-space.
 \begin{equation}
 Z ~\propto~\int\;d^{3nP}\!R\;\exp -\frac{\beta}{P}\sum_{p=1}^P\; \left[
   {\cal V}\left(\vect R\pp\right) 
   + \frac{mP^2}{2\hbar^2\beta^2} {\Delta \vect R\pp}^2 \right].
\label{eqn:nofr}
\end{equation}

where $\beta^{-1}=k_BT$ and $P\in\mathds{N}$ is a large integer, $\Delta \vect R\pp=\vect R\pp-\vect R\ppp$ and $\vect R^{(P+1)}=\vect R^{(1)}$. This partition function is equivalent to that of $n$ (classical) \emph{ring polymers} (one per particle), each composed of $P$ elements (beads) connected by classical springs with spring constant $D=\frac{mP^2}{\hbar^2\beta^2}$. The ensemble of $n$ particles for a given bead is called a replica. The particles within the $p^{\rm th}$ replica interact through the potential ${\cal V}\left(\vect R\pp\right)$, which in our case is the total  electronic energy.

Every replica is a classical representation to the quantum system under study, while the ensemble of the $P$ coupled replica is isomorphic to the quantum system of $n$ atoms. The isomorphism is exploited to compute the solution of the $n$-particle quantum problem via a thermodynamical phase space sampling of the $n$ coupled classical ring polymers. This sampling can be done by means of Monte Carlo techniques or molecular dynamics (MD) simulations, our choice being the latter (using the CPMD code~\cite{cpmd}). Expectation values of all quantum observables which are local in direct space can be expressed via the statistical density of beads, in particular the probability density
\bea
n\nuc(\vect R) &=&\Tr\;\;\left( \left| \vect R\rangle
  \langle\vect R \right| \op{\rho}\T\right)
  \label{eqn:nofr_a}\\
  &\hat{=}& 
    \left\langle\delta^{3n}\left(\vect R-\op{\vect R}\pone\right)
   \right\rangle_{\text{ring polymer}}.
  \label{eqn:nofr_b}
\eea

Properties which are represented in the space of the momenta $\vect{p}=\hbar \vect{k}$, however, require the momentum density
$n\nuc(\vect k)=
\Tr\;\left(\left| \vect k\rangle \langle\vect k \right|\;\op{\rho}\T\right)$,
and thus a modification to the original path-integral formulation. Since $\left| \vect k\right\rangle$ is an eigenstate of the kinetic energy operator $\OP{T}$ and 
$\left\langle\vect R \left| \vect k\right\rangle\right. = \ee^{-i\vect k \vect R}$,
we obtain
\bea
\lefteqn{
n\nuc(\vect k)~=~
 \;\;\ee^{-\betap {\cal T}(\vect k)}
  \int \; d^{3n}\!R\pone\;d^{3n}\!R\ptwo\;\times} \nonumber \\
& &  \;\;\ee^{-i\vect k \left(\vect R\pone -
                      \vect R\ptwo\right)}
  \;\;\rho\left(\vect R\pone,\vect R\ptwo\right).
  \label{eqn:nofp}
\eea
with the off-diagonal density matrix element
$\rho\left(\vect R,\vect R^\prime\right)$.
The latter can be sampled within the classical path-integral isomorphism via a \textsl{open} polymer, where the classical spring between replica $\vect R\pone$ and $\vect R\ptwo$ has been eliminated~\cite{pi-ceperley-rmp}. Using this open path integral formalism at $P=32$ within a gradient-corrected DFT framework~\cite{becke,lyp} 
we have computed the direct- and reciprocal space density distributions $n(\vect r)$ and $n(\vect k)$ of the hydrogen atoms in KDP and DKDP crystals by opening the paths for only these atoms~\footnote{All hydrogen atoms in a cell were represented by open paths neglecting any direct H-H interactions. Since the H-bonds in this system are well-separated this is a good approximation. See ref.~\cite{morrone-pi} for other path-opening strategies in more complicated cases.}. The simulation temperatures were chosen at least about 20K above the respective transition temperatures (140K and 300K), in order to compensate for the known temperature issues in DFT-based MD calculations~\cite{morrone-water}. Our supercell for the paraelectric phases contains four independent phosphate units  on body-centered tetragonal sites, which gives us sufficient statistics (8 protons per cell) but ignores coupling to phonons away from the zone-center. The lattice parameters for each isotope were taken to be the experimentally measured values in the paraelectric phases close to the $T_c$.  For KDP $a=7.4264\AA, c=6.931\AA$, and for DKDP $a=7.459\AA, c=6.957$. Note that the difference in volume is only about 1.3\%. The total simulated time was about 10ps (at a 0.1fs time-step) ensuring good statistics covering both localized and delocalized paths. 

As illustrated in fig.~\ref{OHO}, the KDP crystal contains a two-dimensional network of OH$\cdots$O hydrogen bonds between the phosphate units, orthogonal to the $c$-axis along which the electric polarization is oriented in the ferroelectric phase. We characterize the H-bonds by the oxygen-oxygen distance $2R$  and a hydrogen-localization coordinate
$\delta = R - {\mathbf R_{O_l H}} \cdot {\mathbf R_{O_l O_u }}$
which quantifies the asymmetry of the OH$\cdots$O bond. 
Here, the subscripts $l$($u$) stand for the ``lower'' (``upper'')  oxygen,  ${\mathbf R_{O_l H}}$ refers to the instantaneous vector from the lower oxygen to the hydrogen, and $\mathbf R_{O_l O_u }$ is the unit vector pointing from the lower to the upper oxygen. Each MD step provides a set of $P$ values for $\delta$, whose statistical distribution $n(\delta)$ over the PIMD run defines a hydrogen-localization function (HLF). The HLF represents the projection of the real-space density of the hydrogens along the OH$\cdots$O bond.
A negative (positive) value of $\delta$ implies that the hydrogen is closer to the lower (upper) oxygen O$_l$ (${\rm O}_u$) of the H-bond  (see fig.~\ref{OHO}).

Fig.~\ref{fig:PLF}a shows the HLF for protons in KDP resulting from the PIMD simulations. The broad HLF can result from either quantum delocalization or a classical hopping across a barrier in the H-bond.  However, a conventional Car-Parrinello (CP) simulation at the same conditions showed no hopping at all (see the inset in fig.~\ref{fig:PLF}a) but, instead, resulted in a symmetry-broken ferroelectric state. Hence, in the time-scale of this simulation, a \textit{classical} proton cannot overcome the OH$\cdots$O barrier at 140~K. This implies that the broad HLFs are due to quantum delocalization of the protons in the H-bonds, leading to the symmetric paraelectric phase. 

The computed HLFs of the deuterons in paraelectric DKDP at 300~K is shown in fig.~\ref{fig:PLF}a (bottom left). In contrast to KDP, we see a diversification of the HLF ranging from well-localized off-centered deuterons to considerably delocalized ones. An overall average in them would lead to a symmetric distribution. Unlike in KDP, this symmetry results from \textit{classical} disordering of the deuterons between the two OH$\cdots$O equilibrium sites. A complementary fingerprint of the paraelectric KDP (DKDP), is the \prot (\deut) nuclear momentum distribution $n({\mathbf k})$ along the OH$\cdots$O bond from our linear-polymer path-integral approach. These are shown in 
fig.~\ref{fig:PLF}b along with the corresponding experimental data from neutron Compton scattering experiments~\cite{ReiterKDP}. The good agreement with experiment confirms the \prot-tunneling picture in KDP. The $n({\mathbf k})$ of the deuterons is broader than that of the protons, which corresponds to a tighter localization in real-space. In fact, the experimental $n_{[{}^2 H]}({\mathbf k})$ corresponds to an even higher degree of spatial localization further validating the  \textit{classical} disordering picture.

\begin{figure*}[t!]
\centering
\includegraphics[width=1.0\linewidth,clip=]{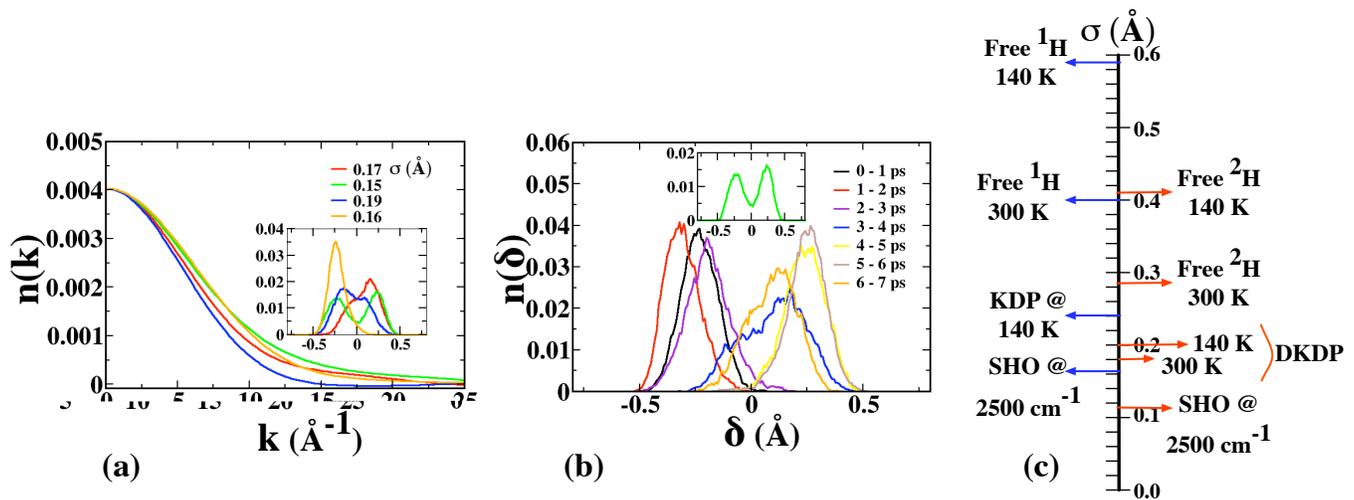}
\caption{(a) Computed momentum and real-space (inset) densities of several deuterons in DKDP at T=300K, and their spreads in \AA~ (inset); (b) Evolution of the $n_{[{}^2H]}(\delta)$ of one specific \deut\ in subsequent 1ps-windows; and (c) the computed spreads ${\langle k^2 \rangle}^{-1}$  of $n(k)$ for \prot\ (left) and \deut\ (right) for a free particle,  the harmonic oscillator and KDP at different temperatures.
\label{fig:nofk}}
\end{figure*}

We have further computed $\sigma=\sqrt{{\langle k^2 \rangle}^{-1}}$ as a measure of the \textit{instantaneous} real-space hydrogen localization (see the  $\sigma$ scale-bar in fig.~\ref{fig:nofk}c). While the deuterons are more localized than the protons, the spread of either species lies between the extremes of a free particle and a simple harmonic oscillator. Interestingly, deuteration at the temperature and volume conditions of the paraelectric KDP simulation resulted in only partial localization (DKDP at 140~K). The localization is enhanced upon increasing the temperature. This enhancement is closely linked to the so-called geometry-effect of isotopic substitution and, in fact, a key feature of the mechanism of the ferroelectric phase transition in DKDP.

The geometry effect manifests itself in experiments as an increased OH$\cdots$O distance ($2R$ in fig.~\ref{OHO}) in DKDP with respect to KDP in the paraelectric phase~\cite{Nelmesrev}. Our PIMD simulations reproduced this experimental trend with an average $2R$ of 2.49~\AA\ in KDP and 2.60~\AA\ in DKDP. Furthermore, we observed a correlation between the instantaneous $2R$ values and the  localization $\delta$ of the hydrogens, shown in fig.~\ref{fig:PLF}a.
 At shorter $2R$ the hydrogen prefers the center of the H-bond while for larger $2R$ the most-probable positions of the hydrogen are off-center. In fact, for very short H-bonds ($2R < 2.45$~\AA) the off-center peaks are absent suggesting the disappearance of a symmetry-broken phase at high pressures~\cite{SamaraKDP}. Thus, the probability of finding the hydrogen at the center of the H-bond is effectively modulated by the acceptor-donor distance $2R$, which determines the tunneling barrier~\cite{Tosatti, Radhousky}.
The actual tunneling regime corresponds to a rather thin range of $2R$ values where the $\delta$ coordinate has a distinctively larger spread about the center. This range (occuring around 2.49~\AA) is wider for KDP than DKDP due to the higher mass of the \deut.
The stabilization via tunnel-splitting in KDP is strong enough to soften of the ferroelectric mode (that is coupled to the $2R$ distances). Thus, the phase transition in KDP can also be thought of as a soft-mode driven transition. This effect is absent in DKDP as the higher mass of the \deut\  diminishes the tunneling probability and hence the tunnel-splitting. 

As the deuterons in the paraelectric phase of DKDP are localized, they must be (classically) disordered over the two OH$\cdots$O sites. The distinction betwen a frustrated antiferroelectric or dynamically disordered character of this state can be made using the individual momentum distributions as well as the corresponding real-space localization functions (see fig.~\ref{fig:nofk}). The real-space picture shows that there are both localized and delocalized deuterons; the momentum distributions, however, all have very similar widths. In some cases, deuterons with a broader real-space distribution (e.g. the green  line in fig.~\ref{fig:nofk}a) exhibit an even broader momentum-space distribution than more localized ones (e.g. the blue line in fig.~\ref{fig:nofk}a). 

This apparent inconsistency is resolved by following the the {}\textsl{temporal evolution} of the real-space density of the green hydrogen, shown in fig.~\ref{fig:nofk}b. Initially, the \deut\ is localized at negative $\delta$ values (black, red, violet lines), but it becomes delocalized after a few picoseconds (blue line) and subsequently localizes at positive $\delta$ (yellow and brown lines). At the end of our run, further delocalization starts (orange line). This plot illustrates that the \deut\ is actually localized most of the time, but jumping from one oxygen site to the other, passing through a delocalized quantum state.

As the \deut\ moves towards the bond-center the corresponding $2R$ also decreases (see the correlation plot in fig.~\ref{fig:PLF}a). This not only reduces the barrier to cross the bond-center but also leads to a  larger real-space width in $n_{[{}^2H]}(\delta)$~\cite{Tosatti}. Thus, the HLF is the widest as the \deut\ passes through the center of the H-bond. Beyond this point $2R$ increases again simultaneously localizing the \deut. Hence, the oscillations in the acceptor-donor distance $2R$, effected by rotational/vibrational modes of the phosphate groups are ultimately responsible for the dynamic disordering of the deuterons in paraelectric DKDP. The amplitudes of these modes at lower temperatures  (in particular around the KDP $T_c$ of 140 K) are not sufficient to enable the hopping process . Thus the ferroelectric phase transition in DKDP occurs via a vibration-assisted hopping of the deuterons across the H-bonds leading to a symmetric paraelectric phase.

In conclusion, we find that the paraelectric phase in KDP is stabilized by a delocalization state (tunneling) of the protons in the OH$\cdots$O hydrogen bonds, whereas the paraelectric state of DKDP results from classical disordering of the deuterons, leading to a symmetrization in a statistical sense. Our results indicate that deuteron hopping is strongly coupled to the vibrations of the oxygens that form the H-bonds (vibration-assisted hopping). This fundamental difference in the mechanism for the ferroelectric phase transition turns out to be responsible for the huge isotope-effect that is observed experimentally.

\bibliography{literatur_daniel}
\bibliographystyle{apsrev}

\end{document}